\documentclass[english,prl,aps,twocolumn,superscriptaddress]{revtex4-2}
\usepackage[T1]{fontenc}
\usepackage[latin9]{inputenc}
\usepackage{amsmath}
\usepackage{amssymb}
\usepackage{graphicx}

\makeatletter
\usepackage{color}
\usepackage[colorlinks,citecolor=blue,linkcolor=blue]{hyperref}

\makeatother

\usepackage{babel}
\begin{document}
\title{Amplifying Frequency Up-Converted Infrared Signals with a Molecular
Optomechanical Cavity}
\author{Fen Zou}
\affiliation{Center for Theoretical Physics \& School of Physics and Optoelectronic
Engineering, Hainan University, Haikou 570228, China}
\affiliation{Beijing Computational Science Research Center, Beijing 100193, China}
\author{Lei Du}
\affiliation{Center for Quantum Sciences and School of Physics, Northeast Normal
University, Changchun 130024, China}
\author{Yong Li}
\email{yongli@hainanu.edu.cn}

\affiliation{Center for Theoretical Physics \& School of Physics and Optoelectronic
Engineering, Hainan University, Haikou 570228, China}
\author{Hui Dong}
\email{hdong@gscaep.ac.cn}

\affiliation{Graduate School of China Academy of Engineering Physics, Beijing 100193,
China}
\date{\today}
\begin{abstract}
Frequency up-conversion, enabled by molecular optomechanical coupling,
has recently emerged as a promising approach for converting infrared
signals into the visible range through quantum coherent conversion
of signals. However, detecting these converted signals poses a significant
challenge due to their inherently weak signal intensity. In this work,
we propose an amplification mechanism capable of enhancing the signal
intensity by a factor of 1000 or more for the frequency up-converted
infrared signal in a molecular optomechanical system. The mechanism
takes advantage of the strong coupling enhancement with molecular
collective mode and Stokes sideband pump. This work demonstrates a
feasible approach for up-converting infrared signals to the visible
range.
\end{abstract}
\maketitle
\emph{Introduction---}The mid- and far-infrared frequency range,
encompassing wavelengths from 2.5 to 500\,$\mu\text{m}$, is a critical
region of the electromagnetic spectrum with significant applications
in many fields, including thermal imaging~\citep{Tonouchi2007Cutting},
quantum sensing~\citep{Kutas2020Terahertz}, microscopy~\citep{Kviatkosvky2020Microscopy,Paterova2020Hyperspectral},
clinical medicine~\citep{Bruyne2018Applications}, and astronomical
surveys~\citep{Ariyoshi2016Terahertz,Roellig2020Mid}. However, detecting
photons within this range presents a significant challenge, as conventional
infrared detectors are sensitive to the thermal noise at the frequency
vicinity of these photons, and require cryogenic temperatures to reduce
this noise. As a result, there is a pressing need for improved detection
technologies that can operate within this frequency range without
the need for cryogenic cooling.

One promising approach is to utilize coherent up-conversion technology~\citep{Heilweil1989Ultrashort,Dougherty1994Dual,Karstad2005Detection,Temporao2006Mid,Tidemand2016Mid,Tseng2018Upconversion,Barh2019Parametric}
to convert lower-frequency infrared (IR) light into the visible or
near-infrared (VIS/NIR) range, which can be detected by using cost-effective
and highly sensitive VIS/NIR cameras. This strategy takes advantage
of the well-developed infrastructures and capabilities of VIS/NIR
cameras, including their high integration and low cost, making it
an attractive option for a wide range of applications. Recently, molecular
optomechanical cavities have emerged as a promising candidate to achieve
coherent frequency up-conversion~\citep{Roelli2020Molecular}. In
such systems, the molecular vibrational motion is bilinearly coupled
to the IR field and optomechanically coupled to the VIS field to be
converted into. The advantages of coupling multiple molecules and
enhancing coupling strength via plasmonic nanocavities result in a
significant enhancement of the detection efficiency~\citep{Chen2021Continuous,Xomalis2021Detecting}
beyond the conventional optomechanical cavity setups~\citep{Tian2010Optical,Dong2012Optomechanical,Hill2012Coherent,ruesink2018Optical,Palomaki2013Coherent,Wang2012Using,Tian2012Adiabatic,Bochmann2013Nanomechanical,Andrews2014Bidirectional,Forsch2020Microwave,Metelmann2014Quantum,Nunnenkamp2014Quantum,Nonreciprocal2015Metelmann,Shen2018Reconfigurablea}.
The question still remains as to whether this system is capable of
detecting a weak IR signal at the few-photon level with enhanced sensitivity.

In this Letter, we propose a scheme to amplify the intensity for the
frequency up-converted IR signal in the molecular optomechanical system
with a blue-detuned pump field, tuned close to the first Stokes sideband
frequency of VIS mode. We show that the coherent IR signal of interest
can be amplified by a factor of 1000 (or more). Such an amplification
mechanism is proved to exist for the blue-detuned pump scheme, while
being absent in the pioneer works with a red-detuned pump field~\citep{Roelli2020Molecular,Chen2021Continuous,Xomalis2021Detecting}.
Additionally we present the necessary stability analyses to show the
parametric regions where the amplification scheme remains stable.
The current theoretical study provides a routine strategy to analyze
the up-converted IR detection, covering both amplification and stability
considerations. And our amplification scheme will enable detecting
the frequency up-converted IR signal of weak intensity.

\begin{figure}
\center \includegraphics[width=0.47\textwidth]{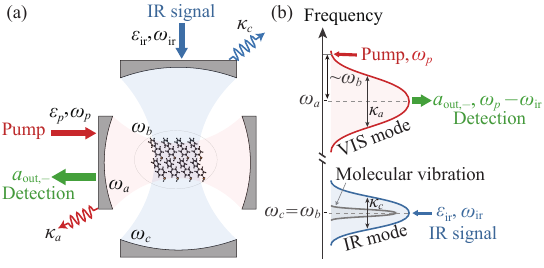} \caption{(a) The molecular optomechanical system consisting of $N$ molecules
(with frequency $\omega_{b}$ of the vibrational mode) coupled to
both the VIS mode (with frequency $\omega_{a}$ and decay rate $\kappa_{a}$)
via the optomechanical interaction and the IR mode (with frequency
$\omega_{c}$ and decay rate $\kappa_{c}$) via the bilinear interaction.
The VIS mode is driven by a pump field with frequency $\omega_{p}$
and amplitude $\varepsilon_{p}$. The IR signal of interest with frequency
$\omega_{\mathrm{ir}}$ and amplitude $\varepsilon_{\mathrm{ir}}$
is incident on the cavity with coupling to the IR mode $c$. (b) Scheme
of frequency up-converted IR signals based on the molecular optomechanical
system. Here the IR signal of interest is near-resonant with the vibrational
frequency of the molecules (as well as the IR mode $c$) and the blue-detuned
pump field is near-resonant with the first Stokes sideband of the
VIS mode, i.e., $\omega_{\mathrm{ir}}\simeq\omega_{b}=\omega_{c}$
and $\omega_{p}\simeq\omega_{a}+\omega_{b}$. The input weak IR signal
with frequency $\omega_{\mathrm{ir}}$ is up-converted as the output
VIS signal ($a_{\text{out},-}$) with frequency $\omega_{p}-\omega_{\mathrm{ir}}$
via the optomechanical interaction between the VIS mode and the molecular
vibration.}
\label{Fig1}
\end{figure}

\emph{Model.---}The molecular optomechanical system consists of $N$
molecules and a cavity supporting both VIS and IR modes, as shown
in Fig.~\ref{Fig1}(a). The cavity modes may consist of plasmonic
modes of nanoparticles~\citep{Chen2021Continuous,Xomalis2021Detecting},
supporting two plasmonic modes with frequencies $\omega_{a}$ in the
VIS region (with annihilation operator $a$) and $\omega_{c}$ in
the IR region (with annihilation operator $c$). Molecules are specifically
chosen to couple with both the VIS and IR modes~\citep{Roelli2020Molecular,Chen2021Continuous,Xomalis2021Detecting}.
A strong pump field in the visible range with frequency $\omega_{p}$
and amplitude $\varepsilon_{p}$ is applied to drive the VIS mode
$a$ in the cavity. The weak IR signal of interest with frequency
$\omega_{\mathrm{ir}}$ and amplitude $\varepsilon_{\mathrm{ir}}$
is incident on the cavity and couples to the IR mode $c$. In the
interaction picture with respect to $H_{0}=\hbar\omega_{p}a^{\dagger}a$,
the Hamiltonian of the system is given as ($\hbar=1$)
\begin{align}
H_{\text{sys}} & =\Delta_{0}a^{\dagger}a+\omega_{c}c^{\dagger}c+\sum_{j=1}^{N}\omega_{b}b_{j}^{\dagger}b_{j}\nonumber \\
 & \quad+\sum_{j=1}^{N}g_{a}a^{\dagger}a(b_{j}^{\dagger}+b_{j})+\sum_{j=1}^{N}g_{c}(c^{\dagger}+c)(b_{j}^{\dagger}+b_{j})\nonumber \\
 & \quad+i(\varepsilon_{p}a^{\dagger}+\varepsilon_{\mathrm{ir}}e^{-i\omega_{\mathrm{ir}}t}c^{\dagger}-\text{H.c.}),\label{eq:HamSys}
\end{align}
where $b_{j}$ ($b_{j}^{\dagger}$) is the annihilation (creation)
operator of the vibrational mode of the $j$th molecule with frequency
$\omega_{b}$. The fourth term describes the optomechanical interaction
between the VIS mode and the molecular vibration with the coupling
strength $g_{a}$ and the fifth term represents the bilinear interaction
between the IR mode and the molecular vibration with the coupling
strength $g_{c}$. The term including $\varepsilon_{p}$ ($\varepsilon_{\mathrm{ir}}$)
describes the coupling of the pump field to the VIS mode (the coupling
of the IR signal to the IR mode). The parameter $\Delta_{0}=\omega_{a}-\omega_{p}$
is the detuning of the VIS mode with respect to the pump field. Without
loss of generality, we assume these parameters ($g_{a}$, $g_{c}$,
$\varepsilon_{p}$, and $\varepsilon_{\mathrm{ir}}$) are real numbers.

By introducing the molecular collective operator $B=\sum_{j=1}^{N}b_{j}/\sqrt{N}$
satisfying $[B,B^{\dagger}]=1$, the Hamiltonian in Eq.~(\ref{eq:HamSys})
is simplified as
\begin{align}
H & =\Delta_{0}a^{\dagger}a+\omega_{c}c^{\dagger}c+\omega_{b}B^{\dagger}B+G_{c}(c^{\dagger}+c)(B^{\dagger}+B)\nonumber \\
 & +G_{a}a^{\dagger}a(B^{\dagger}+B)+i(\varepsilon_{p}a^{\dagger}+\varepsilon_{\mathrm{ir}}e^{-i\omega_{\mathrm{ir}}t}c^{\dagger}-\text{H.c.}),
\end{align}
where $G_{a}=g_{a}\sqrt{N}$ ($G_{c}=g_{c}\sqrt{N}$) is the collective
optomechanical (bilinear) coupling strength. The corresponding quantum
Langevin equations (QLEs) are obtained as
\begin{subequations}
\begin{align}
\dot{a} & =-(i\Delta_{0}+\kappa_{a})a-iG_{a}a(B^{\dagger}+B)+\varepsilon_{p}+\sqrt{2\kappa_{a}}a_{\text{in}},\\
\dot{c} & =-(i\omega_{c}+\kappa_{c})c-iG_{c}(B^{\dagger}+B)+\varepsilon_{\mathrm{ir}}e^{-i\omega_{\mathrm{ir}}t}+\sqrt{2\kappa_{c}}c_{\text{in}},\label{eq:QLEs}\\
\dot{B} & =-(i\omega_{b}+\gamma_{B})B-iG_{a}a^{\dagger}a-iG_{c}(c^{\dagger}+c)+\sqrt{2\gamma_{B}}B_{\text{in}},
\end{align}
\end{subequations}
 where $\kappa_{a}$ ($\kappa_{c}$) and $\gamma_{B}$ are the decay
rates of the VIS (IR) mode and the molecular collective mode respectively.
$a_{\text{in}}$, $c_{\text{in}}$, and $B_{\text{in}}$ are the noise
operators with zero mean values $\langle o_{\text{in}}\rangle=0$
for $o=a,c,B$. The amplitude of the IR signal $\varepsilon_{\mathrm{ir}}$
of interest is much smaller than that of the pump field $\varepsilon_{p}$,
and is therefore treated as a perturbation. The relations between
steady-state mean values of the operators are obtained by neglecting
the term $\varepsilon_{\mathrm{ir}}\exp(-i\omega_{\mathrm{ir}}t)$
in Eq.~(\ref{eq:QLEs}) as $\langle c\rangle_{\text{ss}}=-iG_{c}(\langle B\rangle_{\text{ss}}+\langle B\rangle_{\text{ss}}^{*})/(i\omega_{c}+\kappa_{c})$,
$\langle B\rangle_{\text{ss}}=-i[G_{a}\vert\langle a\rangle_{\text{ss}}\vert^{2}+G_{c}(\langle c\rangle_{\text{ss}}+\langle c\rangle_{\text{ss}}^{*})]/(i\omega_{b}+\gamma_{B})$,
and $\langle a\rangle_{\text{ss}}=\varepsilon_{p}/(i\Delta+\kappa_{a})$.
Here $\Delta=\Delta_{0}+G_{a}(\langle B\rangle_{\text{ss}}^{*}+\langle B\rangle_{\text{ss}})$
represents the effective detuning, modified by the collective optomechanical
interaction. These steady-state mean values are solved self-consistently.

The up-converted IR signal is analyzed via the quantum fluctuation
$\delta o=o-\langle o\rangle_{\text{ss}}$ ($o=a,c,B$) on top of
the steady-state value~\citep{Weis2010Optomechanically}. By keeping
only the first-order term of quantum fluctuation, we obtain the linearized
QLEs as
\begin{align}
\delta\dot{a} & =-(i\Delta+\kappa_{a})\delta a-i\mathcal{G}_{a}(\delta B^{\dagger}+\delta B)+\sqrt{2\kappa_{a}}a_{\text{in}},\nonumber \\
\delta\dot{c} & =-(i\omega_{c}+\kappa_{c})\delta c-iG_{c}(\delta B^{\dagger}+\delta B)+\varepsilon_{\mathrm{ir}}e^{-i\omega_{\mathrm{ir}}t}\nonumber \\
 & \quad+\sqrt{2\kappa_{c}}c_{\text{in}},\nonumber \\
\delta\dot{B} & =-(i\omega_{b}+\gamma_{B})\delta B-i(\mathcal{G}_{a}^{*}\delta a+\mathcal{G}_{a}\delta a^{\dagger})\nonumber \\
 & \quad-iG_{c}(\delta c+\delta c^{\dagger})+\sqrt{2\gamma_{B}}B_{\text{in}},\label{eq:linQLEs}
\end{align}
where $\mathcal{G}_{a}=G_{a}\langle a\rangle_{\text{ss}}$ is the
enhanced collective optomechanical coupling strength due to the strong
pump field. Note that the $\varepsilon_{\mathrm{ir}}\exp(-i\omega_{\mathrm{ir}}t)$
term is now included in Eq.~(\ref{eq:linQLEs}). Here we do not employ
the rotating-wave approximation for either the optomechanical or bilinear
interaction terms.

To solve the linearized QLEs~(\ref{eq:linQLEs}), we use the ansatz
$\langle\delta o\rangle=o_{+}e^{-i\omega_{\mathrm{ir}}t}+o_{-}e^{i\omega_{\mathrm{ir}}t}$
for $o=a,c,B$~\citep{Agarwal2010Electromagnetically,Weis2010Optomechanically,Xu2015Controllable,Zhang2018Loss},
where $o_{+}$ and $o_{-}$ correspond to the values of positive-
and negative-frequency components. With this ansatz, $o_{\pm}$ and
$o_{\pm}^{*}$ are solved analytically with their exact expressions
presented in the Supplementary Material~\citep{SM,Xu2015Optical,Agarwal2012Optomechanical}.

\begin{figure}
\center \includegraphics[width=0.47\textwidth]{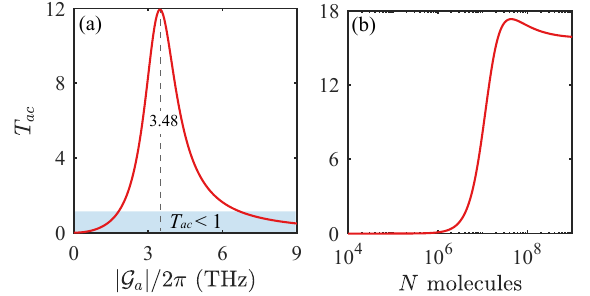} \caption{(a) The conversion efficiency $T_{ac}$ as a function of the enhanced
collective optomechanical coupling strength $\vert\mathcal{G}_{a}\vert$
for $N=10^{7}$. (b) The conversion efficiency $T_{ac}$ as a function
of the number of the molecules $N$ at $g_{a}/2\pi=0.08\,\text{GHz}$.
Here we consider the resonance case $\omega_{\mathrm{ir}}=\omega_{b}=\omega_{c}=2\pi\times30\,\text{THz}$
and other parameters are $\Delta=-\omega_{b}$, $\kappa_{a}/2\pi=30\,\text{THz}$,
$\kappa_{c}/2\pi=0.5\,\text{THz}$, $\gamma_{B}/2\pi=0.16\,\text{THz}$,
$\varepsilon_{p}/2\pi=500\,\text{THz}$, and $g_{c}/2\pi=0.1\,\text{GHz}$.}
\label{Fig2}
\end{figure}

\emph{Amplification.---}The IR signal of interest is amplified when
the frequency of the blue-detuned pump field is tuned close to the
first Stokes sideband of the VIS mode, i.e., $\Delta\simeq-\omega_{b}$.
The up-converted signal is included in the quantum fluctuation $\delta a$
of the VIS mode. The detection is performed on the output field of
the VIS mode, denoted as $a_{\text{out}}=\langle a_{\text{out}}\rangle_{\text{ss}}+\delta a_{\text{out}}$,
where $\langle a_{\text{out}}\rangle_{\text{ss}}$ ($\delta a_{\text{out}}$)
is the steady-state (fluctuation) component of the output field $a_{\text{out}}$.
The mean value of the fluctuation component is rewritten as $\langle\delta a_{\text{out}}\rangle=a_{\text{out},+}e^{-i\omega_{\mathrm{ir}}t}+a_{\text{out},-}e^{i\omega_{\mathrm{ir}}t}$,
where $a_{\text{out},+}$ and $a_{\text{out},-}$ are the first anti-Stokes
and Stokes components of $\langle\delta a_{\text{out}}\rangle$, respectively.
With the input-output relation $a_{\text{out}}+a_{\text{in}}+\varepsilon_{p}/\sqrt{2\kappa_{a}}=\sqrt{2\kappa_{a}}a$~\citep{Gardiner1985Input},
we obtain the output signal as $a_{\text{out},\pm}=\sqrt{2\kappa_{a}}a_{\pm}$.

In the case of the blue-detuned pump field with $\Delta=-\omega_{b}$,
our main focus is on the upconversion at the first Stokes sideband.
For the near-resonant case $\omega_{\mathrm{ir}}\simeq\omega_{b}=\omega_{c}$,
the first Stokes component $a_{-}$ is obtained explicitly as~\citep{SM}
\begin{equation}
a_{-}=\frac{2i\varepsilon_{\mathrm{ir}}\mathcal{G}_{a}G_{c}\Delta(\Delta-\omega_{\text{ir}}+i\kappa_{a})(\Delta-\omega_{\text{ir}}+i\kappa_{c})}{\mathcal{A}(\omega_{\text{ir}})},\label{eq:am}
\end{equation}
where $\mathcal{A}(\omega_{\text{ir}})=\{[\Delta^{2}-(\omega_{\text{ir}}-i\kappa_{c})^{2}][\Delta^{2}-(\omega_{\text{ir}}-i\gamma_{B})^{2}]-4G_{c}^{2}\Delta^{2}\}[\Delta^{2}-(\omega_{\text{ir}}-i\kappa_{a})^{2}]+4\vert\mathcal{G}_{a}\vert^{2}\Delta^{2}[\Delta^{2}-(\omega_{\text{ir}}-i\kappa_{c})^{2}]$.
For the input IR signal $\varepsilon_{\mathrm{ir}}/\sqrt{2\kappa_{c}}$~\citep{SM},
the conversion efficiency $T_{ac}$ at the first Stokes sideband is
obtained as

\begin{equation}
T_{ac}=\vert t_{ac}\vert^{2}=\left\vert \frac{2\sqrt{\kappa_{a}\kappa_{c}}a_{-}}{\varepsilon_{\mathrm{ir}}}\right\vert ^{2},\label{eq:ConversionEff}
\end{equation}
where $t_{ac}\equiv a_{\text{out},-}/\left(\varepsilon_{\mathrm{ir}}/\sqrt{2\kappa_{c}}\right)$
denotes the conversion coefficient from the IR signal to VIS range
at the first Stokes sideband~\citep{Li2017Optical,Zhang2018Loss}.

For simplicity, we first consider the case where the IR signal is
fully resonant with the vibrational frequency of the molecules (as
well as the IR mode), i.e., $\omega_{\mathrm{ir}}=\omega_{b}=\omega_{c}$.
In this case, the conversion coefficient is $t_{ac}=2\sqrt{\kappa_{a}\kappa_{c}}\mathcal{G}_{a}G_{c}/(G_{c}^{2}\kappa_{a}\eta_{c}^{-1}-\vert\mathcal{G}_{a}\vert^{2}\kappa_{c}\eta_{a}^{-1}+\kappa_{a}\kappa_{c}\gamma_{B}\eta_{B})$,
where $\eta_{a,c}=1+i\kappa_{a,c}/(2\Delta)$ and $\eta_{B}=1+i\gamma_{B}/(2\Delta)$~\citep{SM}.

\begin{figure}
\center \includegraphics[width=0.47\textwidth]{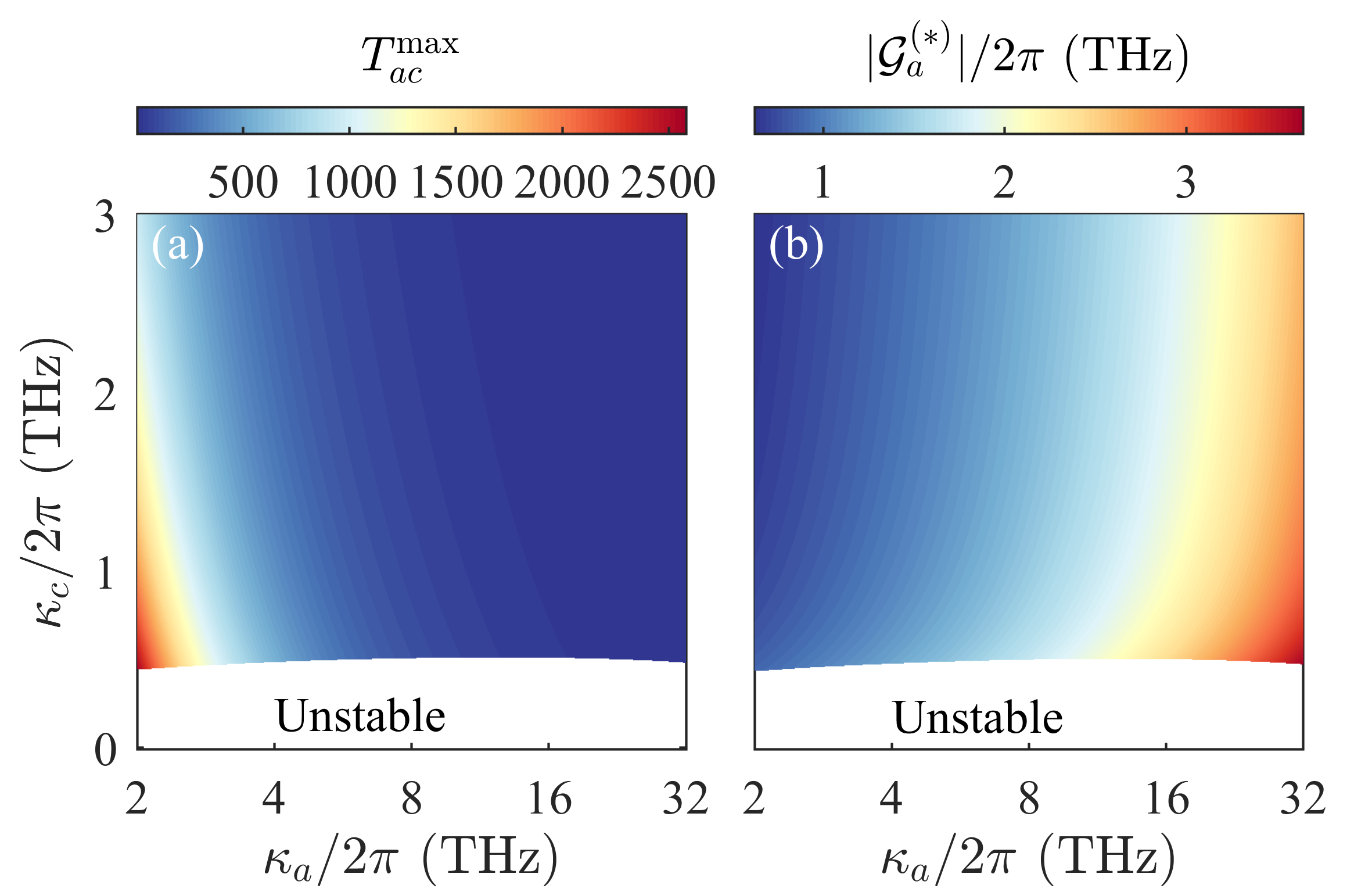} \caption{(a) The maximum conversion efficiency $T_{ac}^{\mathrm{max}}$ as
functions of the decay rates $\kappa_{a}$ and $\kappa_{c}$ at optimal
coupling strength $\vert\mathcal{G}_{a}\vert=\vert\mathcal{G}_{a}^{(*)}\vert\simeq\sqrt{(G_{c}^{2}\vert\eta_{c}^{-1}\vert+\kappa_{c}\gamma_{B}\vert\eta_{B}\vert)\kappa_{a}\vert\eta_{a}\vert/\kappa_{c}}$.
(b) The optimal coupling strength $\vert\mathcal{G}_{a}^{(*)}\vert$
as functions of the decay rates $\kappa_{a}$ and $\kappa_{c}$. Here
$N=10^{7}$ and other parameters are the same as those in Fig.~\ref{Fig2}.}
\label{Fig3}
\end{figure}

Figure~\ref{Fig2}(a) shows the conversion efficiency $T_{ac}$ as
a function of the enhanced collective optomechanical coupling strength
$\vert\mathcal{G}_{a}\vert$ at the first Stokes sideband $\Delta=-\omega_{b}$.
In our numerical simulations, we choose the experimentally feasible
parameters~\citep{shalabney2015Coherenta,long2015Coherenta,Chen2021Continuous,Xomalis2021Detecting,Felix2016Single,Roelli2016Molecular,schmidt2016Quantum,Lombardi2018Pulsed,Pannir2022Driving,esteban2022Molecular,chikkaraddy2022Midinfrared}
as $\omega_{\mathrm{ir}}/2\pi=\omega_{b}/2\pi=\omega_{c}/2\pi=30\,\text{THz}$,
$\kappa_{a}/2\pi=30\,\text{THz}$, $\kappa_{c}/2\pi=0.5\,\text{THz}$,
$\gamma_{B}/2\pi=0.16\,\text{THz}$, $\varepsilon_{p}/2\pi=500\,\text{THz}$,
$g_{c}/2\pi=0.1\,\text{GHz}$, and $N=10^{7}$. One scheme for realizing
the molecular optomechanical system involves an Au nanoparticle inside
a nanogroove etched in a gold film to form the plasmonic cavity. Biphenyl-4-thiol
molecules are chosen to support a prominent vibrational mode that
couples to both the VIS and IR modes in the plasmonic cavity~\citep{Chen2021Continuous}.
The single-photon optomechanical coupling strength between the VIS
mode and the molecular vibration can reach $g_{a}\sim2\pi\times100\,\text{GHz}$~\citep{Roelli2016Molecular}.
For other molecules, such as rhodamine 6G molecule, the single-photon
optomechanical coupling strength $g_{a}$ is within the range $2\pi\times\left(0.006-145\right)\,\text{MHz}$~\citep{schmidt2016Quantum}.
Theoretically, the number of the molecules $N$ in the cavity and
the optomechanical coupling strength $g_{a}$ are affected by the
size of plasmonic cavity. Increasing the cavity size leads to a higher
number of molecules but a decreased coupling strength~\citep{Roelli2020Molecular,Roelli2016Molecular}.
However, the overall effect of amplification is determined by the
enhanced collective coupling strength $\mathcal{G}_{a}$, which can
be further enhanced with the average photon number $\left\langle a\right\rangle _{\text{ss}}.$
Detailed discussions on the impact of the number of the molecules
and the coupling strength are presented in the Supplementary Material~\citep{SM}.

The curve in Fig.\ \ref{Fig2}(a) shows that $T_{ac}$ exceeds unity
under the appropriate coupling strength condition to achieve the amplification
of frequency up-converted IR signal. Moreover, the conversion efficiency
of the IR signal reaches a maximum (i.e., $T_{ac}^{\mathrm{max}}\approx12$)
at optimal coupling strength~\citep{SM} $\vert\mathcal{G}_{a}\vert=\vert\mathcal{G}_{a}^{(*)}\vert\simeq\sqrt{(G_{c}^{2}\vert\eta_{c}^{-1}\vert+\kappa_{c}\gamma_{B}\vert\eta_{B}\vert)\kappa_{a}\vert\eta_{a}\vert/\kappa_{c}}=2\pi\times3.48\,\text{THz}$
for the given parameters. This trend inversion of the conversion efficiency
occurs due to the requirement of matching relationship between two
transfer processes described by the beam-splitting interaction $(G_{c}\delta B^{\dagger}\delta c+h.c.)$
and the two-mode-squeezing one $(\mathcal{G}_{a}\delta a^{\dagger}\delta B^{\dagger}+h.c.)$.
The incoming IR signal is firstly converted into the molecular collective
mode $\delta B$, which is then amplified into the VIS mode $\delta a^{\dagger}$.
For the case with $\omega_{b}=\omega_{c}=-\Delta$, the two-mode-squeezing
coupling Hamiltonian $(\mathcal{G}_{a}\delta a^{\dagger}\delta B^{\dagger}+h.c.)$
can be diagonalized in terms of two new normal (Bogoliubov) modes,
whose eigenvalues deviate largely from $\omega_{b}$ when the coupling
strength is too strong. In this too-strong coupling case, the beam-splitting
coupling $(G_{c}\delta B^{\dagger}\delta c+h.c.)$ can be seen as
the largely-detuned coupling between the IR mode and the two normal
modes. Hence, once the coupling strength $\vert\mathcal{G}_{a}\vert$
is too strong (the two transfer processes are far away from the matching
requirement), the IR signal cannot be converted to the molecular collective
mode and the VIS mode. In Fig.~\ref{Fig2}(b), we illustrate the
dependence of the conversion efficiency $T_{ac}$ of the IR signal
on the number of molecules. The curve shows a non-monotonic behavior
with the highest conversion efficiency occurring at $N\sim10^{7}$.
Additionally, the curve shows a plateau where the conversion efficiency
stays constant when the number of molecules is sufficiently large.

Other key parameters that determine the conversion efficiency of the
IR signal are the decay rates $\kappa_{c}$ and $\kappa_{a}$ of the
IR and VIS modes. In Fig.~\ref{Fig3}(a), we plot the maximum conversion
efficiency $T_{ac}^{\mathrm{max}}$ at the first Stokes sideband as
functions of the decay rates $\kappa_{a}$ and $\kappa_{c}$ at optimal
coupling strength $\vert\mathcal{G}_{a}^{(*)}\vert$. For a fixed
decay rate $\kappa_{c}$ (e.g. $\kappa_{c}/2\pi=1\,\text{THz}$),
$T_{ac}^{\mathrm{max}}$ is significantly improved by decreasing the
decay rate $\kappa_{a}$ of the cavity VIS mode. However, when the
decay rate $\kappa_{a}$ is fixed (e.g. $\kappa_{a}/2\pi=30\,\text{THz}$),
$T_{ac}^{\mathrm{max}}$ increases slowly with the increase of the
decay rate $\kappa_{c}$ of the cavity IR mode. In particular, the
conversion efficiency $T_{ac}$ with a factor around 1000 is achieved
at $\kappa_{a}/2\pi=2\,\text{THz}$ (i.e., $\kappa_{a}/\omega_{b}=1/15$).
These results highlight the importance of controlling the decay rates
of the IR and VIS modes for amplifying frequency up-converted IR signal.
Figure~\ref{Fig3}(b) shows the optimal coupling strength $\vert\mathcal{G}_{a}^{(*)}\vert$
as functions of the decay rates $\kappa_{a}$ and $\kappa_{c}$. The
results show that the optimal coupling strength $\vert\mathcal{G}_{a}^{(*)}\vert$
increases (decreases) as the value of $\kappa_{a}$ ($\kappa_{c}$)
increases.

The molecular optomechanical system could become unstable once the
optomechanical and bilinear coupling strengths are strong enough~\citep{Jiang2018Directional}.
We check the stability of the system with amplification and mark the
unstable region in Fig.~\ref{Fig3}. The stability condition of the
system is given explicitly according to the Routh-Hurwitz criterion\ \citep{DeJesus1987Routh,gradshteyn2014table}.
Mathematically, the system is stable only if the real parts of all
the eigenvalues of the coefficient matrix {[}see Eq.~(S10) in the
Supplementary Material{]} are positive. Physically, the negative real
part of the eigenvalue of this coefficient matrix represents the gain
for the related diagonalized normal mode causing the instability.\textbf{
}The detailed discussion of the stability is presented in the Supplementary
Material~\citep{SM}.

\begin{figure}
\center \includegraphics[width=0.47\textwidth]{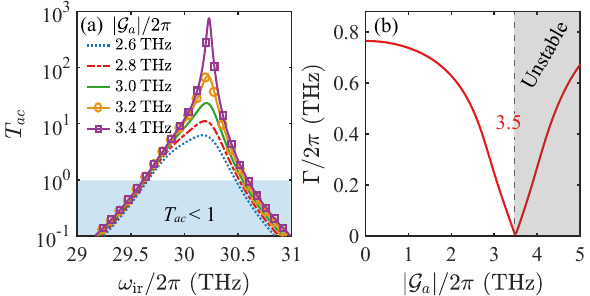} \caption{(a) The conversion efficiency $T_{ac}$ as a function of the frequency
of the IR signal $\omega_{\text{ir}}$ for different values of $\vert\mathcal{G}_{a}\vert$.
(b) The bandwidth $\Gamma$ as a function of the enhanced collective
optomechanical coupling strength $\vert\mathcal{G}_{a}\vert$. Here
$N=10^{7}$ and other parameters are the same as those in Fig.~\ref{Fig2}.
In panel (b), the unstable region is marked with the grey shadow.}
\label{Fig4}
\end{figure}

\emph{Bandwidth of the amplification.---}Another key characteristic
of the up-conversion is the bandwidth of the detection. In the following,
we explore the dependence of the conversion efficiency on the frequency
of the IR signal of interest. For the near-resonant incident IR signal
with $\omega_{\text{ir}}\simeq\omega_{b}=\omega_{c}$, the conversion
efficiency of the IR signal to the VIS range at the first Stokes sideband
is obtained as $T_{ac}=|4\sqrt{\kappa_{a}\kappa_{c}}\mathcal{G}_{a}G_{c}\Delta(\Delta-\omega_{\text{ir}}+i\kappa_{a})(\Delta-\omega_{\text{ir}}+i\kappa_{c})/\mathcal{A}(\omega_{\text{ir}})|^{2}$.
Figure~\ref{Fig4}(a) shows the conversion efficiency $T_{ac}$ as
a function of the frequency $\omega_{\mathrm{ir}}$ of the IR signal
for different enhanced collective optomechanical coupling strengths
$\mathcal{G}_{a}$. As $\vert\mathcal{G}_{a}\vert$ increases, the
maximum conversion efficiency of the IR signal as well as the range
of the amplification increases. For the resonant case $\omega_{\text{ir}}=\omega_{b}=\omega_{c}$,
the maximum conversion efficiency ($T_{ac}^{\mathrm{max}}\approx12$)
is obtained at $\vert\mathcal{G}_{a}\vert/2\pi\approx3.48\,\text{THz}$,
as shown in Fig.~\ref{Fig2}(a). However, for the near-resonant incident
IR signal with $\omega_{\text{ir}}\simeq\omega_{b}=\omega_{c}$, we
find higher amplification than that the resonance case for the frequency
up-conversion. For example, the maximum conversion efficiency $T_{ac}^{\mathrm{max}}$
of the IR signal is approximately $750$ at $\vert\mathcal{G}_{a}\vert/2\pi=3.4\,\text{THz}$,
and the range of the amplification for IR signal is $29.6\,\text{THz}\lesssim\omega_{\text{ir}}/2\pi\lesssim30.6\,\text{THz}$.

The bandwidth $\Gamma$ of the conversion efficiency is obtained by
estimating the full width at half maximum~\citep{Jiang2018Directional,Malz2018Quantum}.
In Fig.~\ref{Fig4}(b), we illustrate the bandwidth $\Gamma$ as
a function of the enhanced collective optomechanical coupling strength
$\vert\mathcal{G}_{a}\vert$, with the unstable region marked with
the gray shadow. The bandwidth $\Gamma$ decreases with the increase
of $\vert\mathcal{G}_{a}\vert$ in the stable region. At $\vert\mathcal{G}_{a}\vert/2\pi\approx3.5\,\text{THz}$,
we observe that the bandwidth is $\Gamma\approx0$ for the near-resonant
case $\omega_{\text{ir}}\simeq\omega_{b}=\omega_{c}$, where the conversion
efficiency of the IR signal diverges. The current figure illustrates
a trade-off between the conversion efficiency and the bandwidth for
choosing the proper coupling strength $\vert\mathcal{G}_{a}\vert$.
The bandwidth of the conversion efficiency is determined by the optomechanical
coupling strength $g_{a}$, the bilinear coupling strength $g_{c}$,
and the decay rate of the molecular vibration $\gamma_{B}$. Detailed
discussions about their effects on the bandwidth are presented in
the Supplementary Material~\citep{SM}. To increase the bandwidth
$\Gamma$, one can use molecules with large decay rate of vibration
$\gamma_{B}$. We also present the equivalent analyses of the conversion
efficiency and the bandwidth with the power spectrum method\ \citep{Clerk2010Introduction,Malz2018Quantum}
in the Supplementary Material~\citep{SM} to confirm the current
results.

\emph{Conclusion and remarks.---}We have proposed an amplification
scheme to increase the sensitivity of detecting coherent IR signal
in the molecular optomechanical systems, where the IR signal is up-converted
into the visible range. In our scheme, the IR signal of interest is
resonant (or near-resonant) with the molecular vibration, and the
blue-detuned pump field, which is near-resonant with the first Stokes
sideband of the VIS mode, is utilized to pump the cavity mode. We
demonstrate the amplification by a factor of several thousands at
the first Stokes sideband of the VIS mode with designed parameters
of the cavity and the molecules and verify the existence of the stability
of scheme. It is worth noting that such an amplification mechanism
is absent for the case of the red-detuned pump field. Detailed discussions
on the conversion efficiency for a red-detuned pump field are presented
in the Supplementary Material. Furthermore, we show the new aspect
with stability analysis, which is important for signal detection in
both blue- and red-detuned regions (see the Supplementary Material).
Our scheme shall provide insight into designing efficient up-conversion
detection of IR signal on the few-photon level.

This work is supported by the National Natural Science Foundation
of China (Grants No.~12088101, No.~12074030, No.~12274107, and
No.~U2230402) and the China Postdoctoral Science Foundation (Grant
No.~2021M700360).


\begin{thebibliography}{54}
\bibitem [{SM()}]{SM}%
  \BibitemOpen
  \href@noop {} {\bibinfo {title} {See {S}upplementary {M}aterial at [url] for
  detailed derivations, which includes {R}efs. [37,38]}}\BibitemShut {NoStop}%
\end{thebibliography}
%

\end{document}